\documentclass[prb,twocolumn,showpacs,superscriptaddress,preprintnumbers,amssymb]{revtex4-2}
\usepackage{graphicx}% Include figure files
\usepackage{color}
\usepackage{dcolumn}% Align table columns on decimal point
\usepackage{bm}% bold math
\usepackage{graphicx}
\usepackage{mathtools}
\usepackage{physics}
\usepackage{hyperref}
\usepackage{slashed}
\usepackage[normalem]{ulem}
\usepackage{tikz}
\usepackage{tikz-cd}
\newcommand{\beq}{\begin{equation}}
\newcommand{\eeq}{\end{equation}}
\newcommand{\beqn}{\begin{eqnarray}}
\newcommand{\eeqn}{\end{eqnarray}}

\newcommand{\ra}{\rightarrow}

\newcommand{\cF}{ {\cal F} }

\newcommand{\cZ}{ {\cal Z} }
\newcommand{\vect}[1]{{\bm{#1}}}

\newcommand{\ii}{\mathrm{i}}

\newcommand{\llangle}{\langle\!\langle}
\newcommand{\rrangle}{\rangle\!\rangle}

\DeclareUnicodeCharacter{2215}{\ensuremath{/}}

\newcommand{\cx}[1]{{\color{black} #1}}

\newcommand{\nmj}[1]{\color{black}}
\definecolor{orange_custom}{rgb}{0.93, 0.47, 0.2}
\newcommand{\as}[1]{{\color{black} #1}}

\begin{document}

\title{Effective Conformal Field Theory generated from Pure and Dephased Chern insulator}

\author{Abhijat Sarma}

\affiliation{Department of Physics, University of California,
Santa Barbara, CA 93106, USA}

\author{Yimu Bao}

\affiliation{Kavli Institute for Theoretical Physics, University of California, Santa Barbara, CA 93106, USA}

\author{Nayan Myerson-Jain}

\affiliation{Department of Physics, University of California,
Santa Barbara, CA 93106, USA}

\author{Thomas Kiely}

\affiliation{Kavli Institute for Theoretical Physics, University of California, Santa Barbara, CA 93106, USA}

\author{Cenke Xu}

\affiliation{Department of Physics, University of California,
Santa Barbara, CA 93106, USA}

\date{\today}

\begin{abstract}

We demonstrate that the fidelity between two states with different Chern numbers $\cZ = \tr\{ \rho \rho' \} $ serves as a generating theory for an effective conformal field theory (CFT) at the $(2+0)d$ temporal interface. $\rho$ can be chosen as a pure trivial insulator, and $\rho'$ can be taken as a pure or dephased Chern insulator density matrix. More specifically, we obtain the following results: (1) through evaluation of the effective central charge, stiffness, and correlation function (the ``strange correlators"), we demonstrate that the fidelity between a trivial insulator and an insulator with Chern number $C=1$ maps to a CFT with effective central charge $c_{\rm eff} = 1$; while the fidelity between two Chern insulators with Chern numbers $C = \pm 1$ maps to a CFT with $c_{\rm eff} = 2$. (2) The density matrix of the Chern insulator becomes a {\it quantum spin Hall insulator} in the doubled Hilbert space, and the dephasing acts as interaction between the two spin species. (3) In the limit of infinite dephasing the Chern insulator becomes a {\it superconductor} in the doubled Hilbert space, featuring the ``strong-weak" U(1) spontaneous symmetry breaking. The analysis based on Laughlin wave function and previous studies of projected wave function of quantum spin Hall insulator suggest this is a power-law superconductor. (4) With increasing strength of dephasing, the amplitude of single particle strange correlator is suppressed, while the Cooper pair strange correlator is enhanced, consistent with the trend of emerging superconductivity. %(5) 

\end{abstract}

\maketitle

\section{Introduction}

In recent years, the phases of open quantum systems, or phases of mixed quantum states of matter have attracted great interest. Like the classification of pure quantum states, a ``phase" should be defined as an equivalence class of states, characterized by the universal behaviors of the class of states. In condensed matter physics, a powerful set of tools capturing universal physics is the formalism based on coarse-graining, including field theory and the renormalization group. Though a complete formalism as such has not yet been developed for open quantum systems, it has been realized that in certain scenarios a field theory description of mixed states of matter can be applicable, in particular for states acted upon by finite-depth quantum channels. For this purpose, the connection to temporal defects in space-time has proven to be very useful. For example, it was shown in Ref.~\cite{altman1} that the effects of weak-measurements and finite-depth decoherence can be mapped to the physics at temporal defects in the Euclidean space-time path-integral formalism, which provides great insights for understanding Luttinger liquid under weak-measurements. Another example of the connection to the temporal defect was made in Ref.~\cite{ashvinshankar,YouXu2013}, which observed that the bulk topology of a symmetry protected topological (SPT) state can manifest at the temporal boundary, especially through the ``strange correlator"~\cite{YouXu2013}, $i.e.$ the correlation function at the temporal interface between the SPT state and a trivial insulator. These connections have facilitated understanding of both quantum critical states and SPT states under weak-measurement and decoherence~\cite{sptdecohere,qibi,altman2,fan2023,wfdecohere,jianmeasure2,aliceameasure,zoumeasure,tarun2024}. 

Within all possible open quantum systems, the topological orders under decoherence are of particular relevance. Quantum information can be stored in topological qubits of a pure topological order, and it can be robust against a certain amount of decoherence. However, strong decoherence can drive a transition beyond which the coherent information stored in topological qubits is lost and irretrievable~\cite{kitaevpreskill,nat1,nat2,natnishimori,leenishimori,wfdecohere,altman2,fan2023}. Among topological orders, the two-dimensional ($2d$) chiral topological order is of particular interest, as it not only hosts anyons that store quantum information, but also features an intrinsic 't Hooft anomaly at the system boundary. This anomaly mandates the presence of gapless modes at the interface between distinct topological states, both at the spatial and temporal interfaces. 

For a chiral topological order, the anyons and 't Hooft anomaly may potentially undergo different changes under decoherence. The goal of this paper is to investigate chiral topological orders under dephasing decoherence. In particular, we would like to diagnose the 't Hooft anomaly of the chiral topological order in the presence of dephasing. Given a bulk wave function of chiral topological order, the potential 't Hooft anomaly can be diagnosed by the strange correlator and related quantities, which has been explored in many past works~\cite{scwierschem1,scwierschem2,scwierschem3,scwierschem4,sczohar1,sczohar2,scscaffidi,scmeng1,scmeng2,scmeng3,scmeng4,scfrank1,scfrank2,scfrank3,scfrank4,scfrank5}. In particular, it was pointed out in Ref.~\cite{cherndecohere} that, when at least one of the two states is pure, the fidelity between the two density matrices maps to the partition function of the conformal field theory (CFT) living at the temporal interface, i.e. slab $\tau = 0, \beta$, in the Euclidean space-time path integral, which will be called the ``fidelity-CFT" hereafter. The relative R\'{e}nyi entropy between the two states maps to the free energy of the fidelity-CFT. If the two states are pure trivial and Chern insulator respectively, theoretical expectation is that the gapless modes at the temporal slab $\tau = 0, \beta$ constitute a $(2+0)d$ gapless Dirac fermion, which is a CFT with central charge $c = 1$. When the Chern insulator is under decoherence, the Kraus operators of the decoherence channel map to the interactions of the Dirac fermion. 

In this work, we consider the fidelity and 2nd relative R\'{e}nyi entropy between two states with different Chern numbers, for example, a pure trivial insulator density matrix $\rho_{tr}$, and a dephased Chern insulator $\rho^g_c$ with dephasing strength $g$. Since the trivial insulator is pure, the fidelity $\cZ$ and relative R\'{e}nyi entropy $\cF$ take a simplified form \beqn \cZ = \tr\{ \rho_{tr} \rho^g_c  \} , \ \ \ \cF = - \ln \cZ. \eeqn %\yimu{$\cF$ is technically not the relative entropy.}
Since we expect these quantities to map to the partition function and free energy of the fidelity-CFT, we would like to extract the quantities of interest for the fidelity-CFT, including the effective central charge, phase stiffness and the scaling dimension of single-particle and composite operators, for different strengths of dephasing.

\section{Zero dephasing}

\subsection{Effective central charge $c_{\rm eff}$}

Topological insulators in an open environment and out of equilibrium have garnered great interest~\cite{cooper,cooper2,cooper3,opentopo,opentopo2,opentopo3,opentopo4,opentopo5,opentopo6,tarun2024}. In this work we will investigate the pure or dephased Chern insulator using the strange correlator and related quantities, such as fidelity and relative R\'{e}nyi entropy. 
When studying a density matrix, it is often convenient to use the doubled Hilbert space representation of the density matrix~\cite{choi,choi2}. For example, in the doubled space the dephased Chern insulator density matrix maps to a pure state \begin{gather} |\rho^g_c \rrangle = \prod_i e^{g \sigma^z_{i,1} \sigma^z_{i,2}} \left( |\Psi_1\rangle \otimes |\Psi_2\rangle  \right), \nonumber \\ \cZ = \tr\{ \rho_{tr} \rho^g_c \} = \llangle \rho_{tr} | \rho^g_c \rrangle. \end{gather} Here $|\Psi_1\rangle$ is a Chern insulator state with Chern number $C = +1$, and $|\Psi_2\rangle$ is the complex conjugate of $|\Psi_1\rangle$, which is a Chern insulator with Chern number $C = -1$. Hence in the doubled space a Chern insulator density matrix maps to a {\it quantum spin Hall insulator}. The dephasing acts like an attractive interaction between the two ``spin" flavors, with $\sigma^z_{i,a} = 2 n_{i,a} - 1$. We would like to extract the quantities of interest for the effective CFT, including the central charge, the phase stiffness, and the scaling dimension of single-particle and Cooper-pair operators. For later use, we also define a rescaled dephasing strength $\alpha$ by $\cos(\alpha) = e^{-g/2}$, so $\alpha=\pi/2$ corresponds to the infinite dephasing limit.

At zero dephasing $g = 0$, the fidelity and relative R\'{e}nyi entropy between a trivial insulator and a Chern insulator (labeled the ``trivial-Chern" quantities) with Chern number $C = +1$ map to the partition function and the free energy of a free Dirac fermion CFT, with central charge $c = 1$. The effective central charge of the fidelity-CFT can in principle be extracted based on the finite-size scaling of the relative R\'{e}nyi entropy, based on the finite size scaling of the free energy of a standard $2d$ CFT~\cite{cardy1986}: \beqn \frac{\cF(L_x, L_y)}{L_y} = f_0 L_x - \frac{\pi c_{\rm eff}}{6 L_x}. \eeqn Here $\cF(L_x, L_y)$ is the relative R\'{e}nyi entropy defined on a $2d$ lattice with rectangular geometry and $L_y \gg L_x$. 
%This calculation can in principle be conducted numerically in the doubled space for any dephasing strength $g$. 

\begin{center}
\begin{figure}[h]
\includegraphics[width=0.48\textwidth]{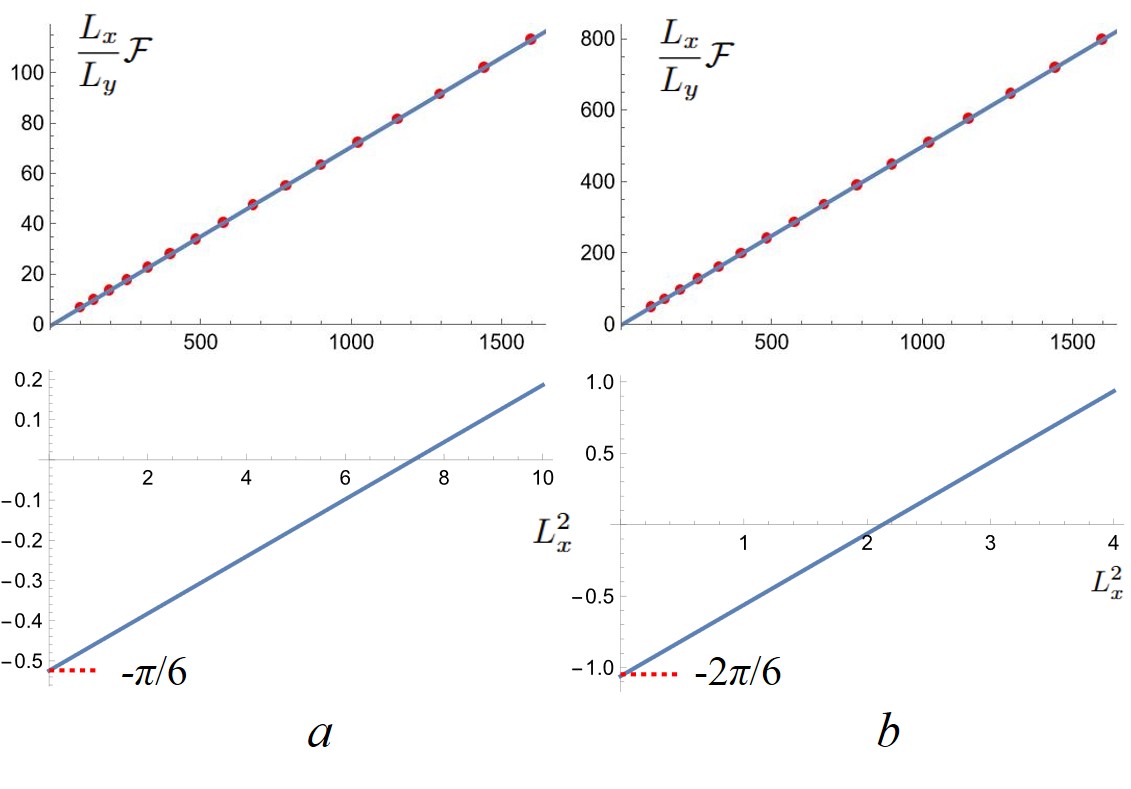}
\caption{($a$) We extract the effective central charge $c_{{\rm eff}}$ for the ``trivial-Chern" fidelity CFT. We fix $L_y = 100$, and plot the computed $\frac{L_x}{L_y}\cF$ vs. $L_x^2$. The intercept is at $ \sim - 0.524$, consistent with the theoretical value $ - \pi/6 \sim - 0.524$ (zoomed-in plot in the lower panel). ($b$) The effective central charge $c_{{\rm eff}}$ for the ``Chern-Chern" fidelity-CFT, with Chern numbers $+1$ and $-1$. %Again, $L_y$ is fixed at $100$, and $\frac{L_x}{L_y}\cF$ is plotted v.s. $L_x^2$. 
The intercept is at $ \sim - 1.06$, also consistent with the theoretical value $ - 2 \pi/6 \sim - 1.05$. } \label{ceff}
\end{figure}
\end{center}

\begin{center}
\begin{figure}[h]
\includegraphics[width=0.37\textwidth]{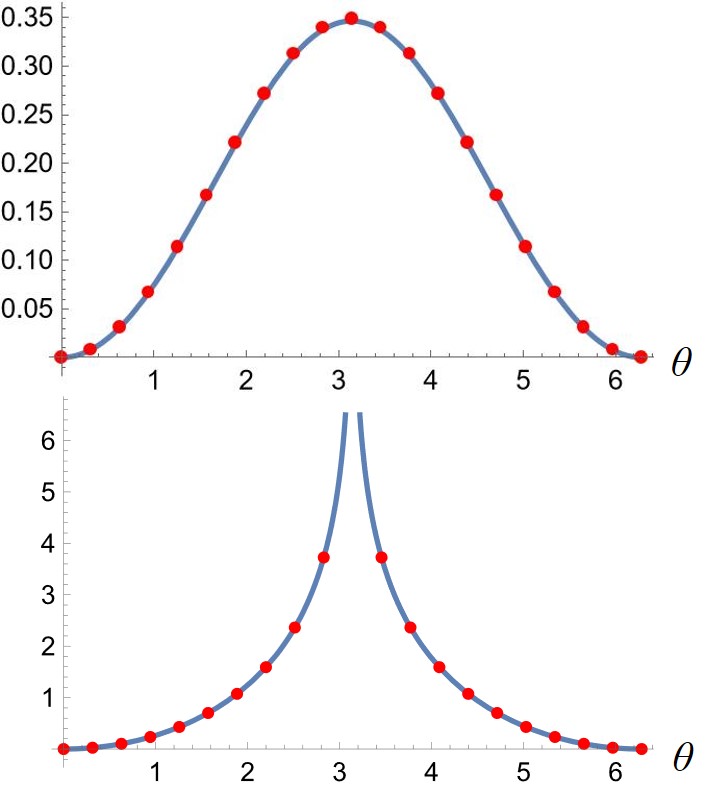}
\caption{The comparison between the trivial-Chern relative R\'{e}nyi entropy (data points), and the free energy of the compact boson after considering twisted sectors based on Eq.~\ref{AA},~\ref{AP} (the solid curves), as a function of extra twisted boundary condition $\theta$. The upper and lower panels are results with ${\rm A/A}$ and ${\rm A/P}$ boundary conditions at $\theta = 0$. } \label{theta}
\end{figure}
\end{center}

%\begin{center}
%\begin{figure}[h]
%\includegraphics[width=0.4\textwidth]{ceff2.jpg}
%\caption{ We extract the effective central charge $c_{{\rm eff}}$ for the fidelity between two pure Chern insulators with Chern number $+1$ and $-1$. Again, $L_y$ is fixed at $100$, and $\frac{L_x}{L_y}\cF$ is plotted v.s. $L_x^2$. The intercept is at $ \sim - 1.06$, also consistent with the theoretical value $ - 2 \pi/6 \sim - 1.05$. } \label{ceff2}
%\end{figure}
%\end{center}
When both the trivial and Chern insulators possess extra spatial discrete symmetries such as inversion, the fidelity-CFT at the temporal interface is unitary. The Hermiticity of the density matrix demands that the fidelity-CFT be invariant under $\psi_L \ra \psi_R^\dagger$, $\psi_R \ra \psi_L^\dagger$ and $\ii \ra - \ii$, where $\psi_{L,R}$ are the left and right moving fermion modes at the temporal interface. This alone does not ensure a unitary fidelity-CFT, as terms such as $\ii \psi_L^\dagger \partial_x \psi_L - \ii \psi_R^\dagger \partial_x \psi_R$ and $\psi_L^\dagger \partial_x \psi_L + \psi_R^\dagger \partial_x \psi_R$ are both allowed in the effective Hamiltonian of the CFT. But if there is an extra inversion symmetry: $x \ra -x$, $\psi_L \leftrightarrow \psi_R$, the second term is excluded, and the fidelity-CFT remains unitary. 

Based on these observations, we use the model for Chern insulator on the square lattice introduced in Ref.~\cite{QWZ}: \beqn H(k_x, k_y) &=& \sin(k_x) \sigma^x + \sin(k_y) \sigma^y \cr\cr &+& (\gamma - \cos(k_x) - \cos(k_y)) \sigma^z. \eeqn The Chern number for the model is $C = +1$ for $ 0< \gamma < 2$, $C = -1 $ for $ -2 < \gamma < 0$, and $C = 0$ for $|\gamma|>2$. The advantage of this model is that, it has plenty of discrete symmetries, which ensures the unitarity of the fidelity-CFT. 

%We first discuss physics at zero dephasing $g = 0$. 
To extract the effective central charge $c_{\rm eff}$, we compute the fidelity and relative R\'{e}nyi entropy with large $L_y$, and finite $L_x$. In the zero dephasing limit, the fidelity is a simple product in the momentum space: \beqn \cZ = \prod_{\vect{k}}|\langle \vect{k} _{tr} | \vect{k}_c \rangle|^2. \label{fid2} \eeqn $|\vect{k}_c\rangle$ and $|\vect{k}_{tr}\rangle$ are the Bloch wave functions of the Chern and trivial insulators, respectively. To avoid singularity in $\cF$ we take anti-periodic boundary condition along at least one of the two directions. We then plot $\frac{L_x}{L_y}\cF$ versus $L_x^2$, the intercept is supposed to be $- \pi c_{{\rm eff} }/6$. 

We first compute the fidelity and relative R\'{e}nyi entropy between two states that are the ground states of $H(\vect{k})$ with $\gamma = 5/2$ and $\gamma = 3/2$, i.e. it is the fidelity between a trivial insulator and Chern insulator with $C = +1$, labeled the ``trivial-Chern" fidelity. The relative R\'{e}nyi entropy in this case is supposed to map to the free energy of a CFT with central charge $c = 1$. The intercept of the plot of $\frac{L_x}{L_y}\cF$ vs. $L_x^2$ is $- 0.524$ (Fig.~\ref{ceff}$a$), consistent with the theoretical expected value $- \pi/6 \sim - 0.524$. We also compute the fidelity and relative R\'{e}nyi entropy between two states with parameters $\gamma = 1$ and $-1$, i.e. the ``Chern-Chern" fidelity, we obtain intercept $- 1.06$ (Fig.~\ref{ceff}$b$), also consistent with the theoretical value $- 2 \pi/6 \sim - 1.05$. 

\subsection{Twisted boundary conditions}

The $(1+1)d$ free Dirac fermion is dual to a compact boson with Luttinger parameter $K = 1$: \beqn S_B = \int d^2x \ \frac{K}{2\pi} (\partial_\mu \Theta)^2. \eeqn To further confirm the existence of the fidelity-CFT, we can compare the ``trivial-Chern" relative R\'{e}nyi entropy, with the free energy of the compact boson. The exact relation between the fermion and compact boson partition functions needs some care. More precisely, an exact relation can be made only when we consider both anti-periodic (A) and periodic (P) boundary conditions along both directions of the fermion theory, and the compact boson with twisted sectors in both directions~\cite{jishao}. When we take the anti-periodic boundary condition for both directions of the fermion, the trivial-Chern fidelity is expected to map to \beqn \cZ_{\rm A/A}  = \frac{1}{2} (Z^B_{00} + Z^B_{10} + Z^B_{01} - Z^B_{11} ), \label{AA} \eeqn where $0$ and $1$ means whether there is a $Z_2$ twisting of bosons along the corresponding direction. Here the $Z_2$ twisting means we sum over half-integer rather than integer number of windings of the compact boson along that direction. When we take anti-periodic boundary along one direction, and periodic along the other direction, the relation is \beqn \cZ_{\rm A/P}  = \frac{1}{2} (Z^B_{00} + Z^B_{10} - Z^B_{01} + Z^B_{11} ). \label{AP} \eeqn 

The relations Eq.~\ref{AA} and Eq.~\ref{AP} are supposed to be valid with extra twisting boundary condition $\theta$ too. We compute the fidelity and relative R\'{e}nyi entropy of fermion states with a twisted boundary condition: \beqn \cZ(\theta) = \tr\{ \rho_{tr}(\theta) \rho_c(\theta)  \}. \eeqn The $\rho_c(\theta) = |\Psi(\theta)\rangle \langle \Psi(\theta)| $ is the density matrix of the ground state of a Chern insulator model with an extra twisted boundary condition $\Psi(x+L) = \eta e^{\ii \theta}\Psi(x)$, where $\eta = \pm 1$ depends on whether we take periodic or anti-periodic boundary condition before twisting. To implement the twisted boundary condition for the relative R\'{e}nyi entropy, we replace $k_x$ in Eq.~\ref{fid2} by $k_x + \theta/L$, and compute $\cF(\theta)$. The computed trivial-Chern relative R\'{e}nyi entropy $\cF_{\rm A/A}(\theta)$ and the compact boson partition function with twisted sectors at $K = 1$ are plotted in Fig.~\ref{theta}, with an excellent agreement. 

%We can define the effective stiffness of the fidelity-CFT as: \beqn \rho_s = - \partial^2_{\theta} \ln \cZ(\theta). \eeqn The $\rho_s$ is independent of the system size. In the doubled space, the density matrix becomes a QSH insulator, and the stiffness $\rho_s$ corresponds to the spin stiffness of the QSH insulator, as the two spin flavors of the state in the doubled space have opposite twisted boundary conditions. At $g = 0$ the extracted $\rho_s$ are \beqn \rho_{s,{\rm A/A}} \sim 0.08, \ \ \ \rho_{s,{\rm A/P}} \sim 0.254. \eeqn These are consistent with the stiffness of the compact boson theory, after summing over different twisted sectors according to Eq.~\ref{AA}, \ref{AP}. 

We can define the effective stiffness of the fidelity-CFT as the following: \beqn \rho_s = - \frac{1}{2}\frac{\partial^2}{\partial \theta^2} \eval_{\theta = 0} \ln \frac{\cZ(\theta)}{\cZ(0)}. \eeqn The stiffness $\rho_s$ is independent of the system size. In the doubled space, the density matrix becomes a QSH insulator, and the stiffness $\rho_s$ corresponds to the spin stiffness of the QSH insulator, as the two spin flavors of the state in the doubled space have opposite twisted boundary conditions. At $g = 0$ the extracted $\rho_s$ are
\beqn
\rho_{s,{\rm A/A}} \sim 0.08 \ \ \ \rho_{s,{\rm A/P}} \sim 0.254
\eeqn
These are consistent with the stiffness of the compact boson theory, after summing over different twisted sectors according to Eq.~\ref{AA}, \ref{AP}. They are also consistent with the stiffness of actual $(1+1)d$ Dirac fermions with corresponding boundary conditions.
%\beqn
%\rho_{s,{\rm A/A}} &=& -2 \sum_{n = 1}^\infty \frac{(-1)^n n q^{n/2}}{1-q^n}  \cr\cr
%\rho_{s,{\rm A/P}} &=& \frac{1}{4} - 2 \sum_{n = 1}^\infty  \frac{(-1)^n n q^n}{1-q^n}
%\eeqn
%where $q = e^{-2 \pi}$ on the torus where both cycles have equal length. 

If we consider the Chern-Chern fidelity-CFT between two states with Chern numbers $+1$ and $-1$, the extracted stiffness is twice the values above, for both A/A and A/P boundary conditions. If we compute relative R\'{e}nyi entropy between two states with the same Chern number, the extracted stiffness vanishes with increasing system size, as the fidelity-CFT would be trivial.

%\beqn \rho_s(\alpha)/\rho_s(0) \eeqn

%an actual $(1+1)d$ Dirac fermion, whose partition function is \beqn Z^{\rm Dirac}(\theta) = {\rm Det} \left( \slashed{D}(\theta) \right), \eeqn where $\slashed{D}$ is the Dirac operator, as well as 

\begin{center}
\begin{figure}[h]
\includegraphics[width=0.4\textwidth]{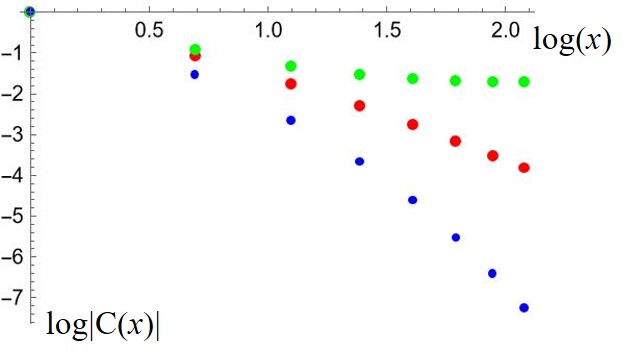}
\caption{The normalized single particle strange correlator $C_\psi^s(\mathbf{x})$ in the zero dephasing limit with A/P (green) and A/A (red) boundary conditions, on a $20 \times 8$ lattice. On a finite system the A/A strange correlator better captures the power-law scaling in the thermodynamic limit. The difference between the A/A and A/P strange correlators is due to the difference between the finite minimum momentum $k_y$. But both decay significantly slower compared to the ordinary single particle Green's function (blue). The boundary condition affects the Green's function very weakly. } \label{SC0}
\end{figure}
\end{center}

\subsection{Strange correlators}

For a $(1+1)d$ free Dirac fermion CFT, the single fermion and Cooper pair correlation both decay with a power-law, which leads to the following predictions for the strange correlators: \beqn C^s_\psi (\vect{x}) &=& \tr\{ \rho_{tr} \psi^\dagger(\vect{0}) \psi(\vect{x}) \rho_c  \} / \cZ \cr\cr &=& \frac{\llangle \rho_{tr}| \psi^\dagger_1(\vect{0}) \psi_1(\vect{x}) | \rho_c \rrangle}{\llangle \rho_{tr}| \rho_c \rrangle}  \sim \frac{1}{|\vect{x}|^{2\Delta_\psi}}, \cr\cr C^s_{\rm cp} (\vect{x}) &=& \tr\{ \rho_{tr} \psi^\dagger(\vect{0}) \psi(\vect{x}) \rho_c \psi(\vect{0}) \psi^\dagger(\vect{x}) \} / \cZ \cr\cr &=& \frac{\llangle \rho_{tr}| \left( \psi_1\psi_2 \right)^\dagger_{\vect{0}} ( \psi_1 \psi_2 )_{\vect{x}} | \rho_c \rrangle} {\llangle \rho_{tr}| \rho_c \rrangle}  \sim \frac{1}{|\vect{x}|^{2\Delta_{\rm cp}}}. \label{corr_eqns} 
%\cr\cr C^s_{spin} (\vect{x}) &=& 1\tr\{ \rho_{in} \psi(\vect{0}) \psi^\dagger(\vect{x}) \rho_c \psi(\vect{0}) \psi^\dagger(\vect{x}) \} \cr\cr &=& \llangle \rho_{in}| ( \psi_1\psi^\dagger_2 )_{\vect{0}} ( \psi^\dagger_1 \psi_2 )_{\vect{x}} | \rho^d_c \rrangle  \sim \frac{1}{|\vect{x}|^2}. 
\eeqn In the free fermion case, $\Delta_{\psi} = 1/2$, and  $\Delta_{\rm cp} = 1$. In contrast, the ordinary single particle Green's function and the Cooper pair correlation in both $\rho_{tr}$ and $\rho_c$ should rapidly decay exponentially because both states are insulators. 

The free-fermion numerics suggests that the correlators defined above crossover to the correct scaling dimension predicted by our field theory with large enough system size, independent of boundary condition. For nonzero dephasing, we evaluate the correlators using determinant Quantum Monte Carlo, and thus we are limited to finite system sizes and must mitigate the finite size effects through appropriate choice of boundary condition. The normalized {\it zero dephasing} single particle strange correlators with A/P and A/A boundary conditions on a finite $20 \times 8$ lattice are plotted in Fig.~\ref{SC0}, together with the ordinary single particle Green's function of the Chern insulator along the same direction. The Green's function is short ranged, but the strange correlators are significantly enhanced, as was expected theoretically. The A/P strange correlator is further enhanced compared with A/A, since for a finite system size the A/A strange correlator is limited with the finite smallest momentum $k_y$. We find that at finite system size, the A/A strange correlator more clearly features the power-law behavior that is consistent with the results in the thermodynamic limit. For the rest of the paper we will always take the A/A boundary conditions for the calculation of the strange correlators. \cx{Note that the free fermion numerics can be performed with much larger system size, and the strange correlator with the A/A boundary condition does feature the correct scaling $1/x^{2\Delta_\psi}$ with $\Delta_\psi = 1/2$ with sufficiently large system. }

\section{ Infinite dephasing $g = + \infty$}

At infinite dephasing $g = + \infty$, the dephasing acts as a ``Gutzwiller projection" in the doubled space, as was noticed in recent work~\cite{choiSL}. The dephased Chern insulator in the doubled space becomes a Gutzwiller projected QSH insulator, i.e. \beqn |\rho_c^\infty\rrangle \sim \prod_i \hat{P} (n_{i,1} = n_{i,2})  \left( |\Psi_1\rangle \otimes |\Psi_2\rangle  \right). \eeqn \cx{Field theory argument suggests that the projected QSH wave function would have a long range superconducting order~\cite{ransu2,yewen1}. But numerics indicates that the state is a power-law superconductor~\cite{ransu2},} with algebraic Cooper pair correlation function, and the scaling dimension of the Cooper pair operator is $1/2$, i.e. \beqn \llangle \rho_c^\infty | \left( \psi_1\psi_2 \right)_{\vect{0}} ( \psi^\dagger_1 \psi^\dagger_2 )_{\vect{x}} |\rho_c^\infty \rrangle \sim \frac{1}{|\vect{x}|}. \label{projectcp} \eeqn

\cx{This power-law superconducting order} can be understood using the real space Laughlin state approximation of the Chern insulator. When represented as the Laughlin state, the Chern insulator wave function reads \beqn \Psi_c(z_i) \sim \prod_{i < j} (z_i - z_j) e^{\sum_i - |z_i|^2}. \eeqn Then in the form of Laughlin wave function, leaving the exponential term implicit, the doubled state of the Chern insulator in the $g \ra \infty$ limit becomes \beqn \Psi^{\infty}_c(z_i) &\sim&  \prod_{i < j} P_{(z_i = w_i)} (z_i - z_j) (w^\ast_i - w^\ast_j) \cr\cr &\sim& \prod_{i < j} |z_i - z_j|^{2}. \label{laughlinSF} \eeqn This is the wave function of a power-law superfluid (superconductor). The Cooper pair correlation function of $\Psi^{\infty}_c(z_i)$ can be evaluated with the Coulomb gas approximation, and it precisely leads to the behavior Eq.~\ref{projectcp}~\cite{girvin}. 

\begin{center}
\begin{figure}
\includegraphics[width=0.38\textwidth]{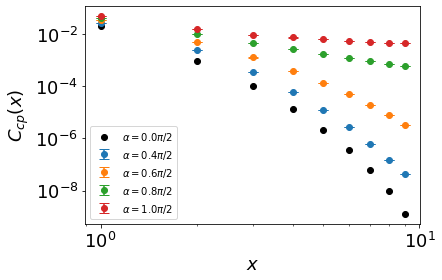}
\includegraphics[width=0.38\textwidth]{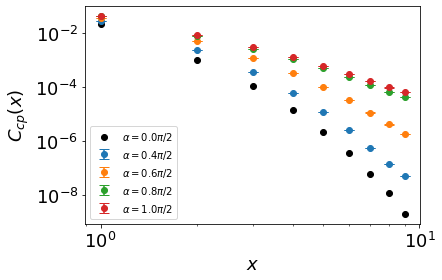}
\caption{
The ordinary Cooper pair correlation in the doubled space (the Renyi-2 correlator) as a function of dephasing strength, on $20 \times 8$ lattice, with periodic (top) and antiperiodic (bottom) boundary conditions along both directions.
%\yimu{with anti-periodic boundary conditions?}. 
The projected Laughlin wave function suggests a $\frac{1}{|\vect{x}|}$ power law emerges at $g \ra \infty$. The correlation with A/A and P/P boundary conditions decay faster and slower respectively compared with $\frac{1}{|\vect{x}|}$ at long distance in the plot. } \label{coopercorre}
\end{figure}
\end{center}

The ordinary Cooper pair correlation function in the doubled space as a function of dephasing strength is shown in Fig.~\ref{coopercorre}. The Cooper pair corelation function is strongly enhanced with an increasing dephasing strength. At $g \ra \infty$, the decay of correlation depends on the boundary conditions in our finite system size: the A/A boundary condition leads to a power-law decay faster than $1/|\vect{x}|$, while the correlation with P/P boundary decays slower than $1/|\vect{x}|$ at long distance in the plot. 
%, which is consistent with a potential power-law superconductor. 
The ordinary correlation function of the doubled space is actually the Renyi-2 correlator $\tr\{ \rho_c^\infty c_{\vect{0}} c^\dagger_{\vect{x}} \rho_c^\infty c_{\vect{x}} c^\dagger_{\vect{0}}  \} $. The emergence of a long-range correlation of the Renyi-2 correlator implies the U(1) strong-weak spontaneous symmetry breaking, which is a subject that has attracted great interest in recent years~\cite{wfdecohere,biswssb,biswssb2,Ogunnaike_2023,huangswssb,yizhiswssb}. 

The wave function in the limit of $g \ra \infty$ is just a bosonic wave function for superfluid. Ref.~\cite{kane1991} pointed out that Eq.~\ref{laughlinSF} is the ground state of bosons interacting through a long range interaction. 
%We can write down a field theory Hamiltonian for the bosons with long range interaction: 
%\beqn \cH = \sum_{\vect{q}} \tilde{K} |\vect{q}|^2 |\phi_\vect{q}|^2 + \frac{u}{\tilde{K}|\vect{q}|^2} |n_\vect{q}|^2, \label{bosonH}  \eeqn where $n_{\vect{q}}$ and $\phi_{\vect{q}}$ are the density and phase fluctuations of the boson, and 
Based on this picture, a purely bosonic ground state wave-functional can be derived: 
%The ground state wave functional of Eq.~\ref{bosonH} can also be written as 
\beqn |\Psi_c^\infty\rangle \sim \int D[\phi] \ e^{ - \int d^2x \frac{1}{8 \pi} (\nabla \phi)^2 } | \phi(x)\rangle, \eeqn the boson (Cooper pair) creation/annhilation operators can be represented as $e^{\pm \ii \phi}$, and $| \phi(x)\rangle$ is the configurational basis of $\phi(x)$. The ordinary Cooper pair correlation of $|\Psi^\infty_c\rangle$ is given by: \beqn  C_{\rm cp}(\vect{x}) &\sim& \langle \Psi^\infty_c | e^{\ii \phi_{\vect{0}}} e^{- \ii \phi_{\vect{x}}} |\Psi^{\infty}_{\rm c} \rangle \cr\cr &\sim& \int D[\phi] \ e^{\ii \phi_{\vect{0}} - \ii \phi_{\vect{x}}} e^{ - \int d^2x \frac{1}{4 \pi} (\nabla \phi)^2 } \cr\cr &\sim& \frac{1}{|\vect{x}|}, \eeqn as was expected from analysis based on the Laughlin wave function. 

In the configurational basis of $\phi(x)$, a bosonic Mott insulator wave function can be written as $\int D[\phi] | \phi(x)\rangle $, which is an equal weight superposition of all configurations of $\phi(x)$, capturing the strong phase fluctuation of the bosonic Mott insulator. Then Cooper pair strange correlation function becomes the path integral \beqn C^s_{\rm cp}(\vect{x}) &\sim& \langle \Psi_{\rm MI} | e^{\ii \phi_{\vect{0}}} e^{- \ii \phi_{\vect{x}}} |\Psi^{\infty}_{\rm c} \rangle \cr\cr &\sim& \int D[\phi] \ e^{\ii \phi_{\vect{0}} - \ii \phi_{\vect{x}}} e^{ - \int d^2x \frac{1}{8 \pi} (\nabla \phi)^2 } \cr\cr &\sim& \frac{1}{|\vect{x}|^{2}}. \eeqn This is the same scaling as the pure trivial-Chern Cooper pair strange correlator.

%The scaling dimension of the bosons is proportional. 

%In the infinite dephasing limit, the Cooper pair strange correlator can be inferred from the behavior of the ordinary correlation function.

\begin{center}
\begin{figure}[h]
\includegraphics[width=0.38\textwidth]{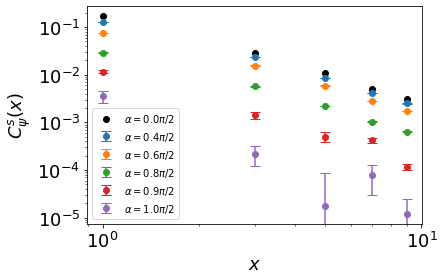}
\includegraphics[width=0.39\textwidth]{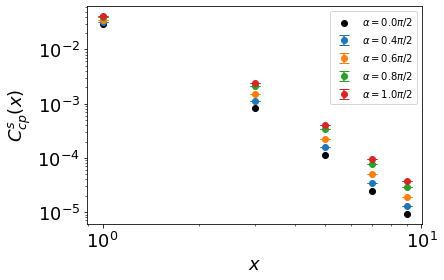}
\caption{The single particle and Cooper pair trivial-Chern strange correlators for different dephasing strengths measured on a $20 \times 8$ lattice. The Cooper pair correlation is enhanced with dephasing, while the single particle strange correlator is suppressed. These are consistent with the power-law superconductivity emerging in the infinite dephasing limit. The last few data points for the single particle strange correlator with $\alpha = \pi/2$ are too small to be observed reliably without more sampling. } \label{SC}
\end{figure}
\end{center}

The single particle trivial-Chern strange correlator should be strongly suppressed in the limit $g = \infty$, as the wave function in this limit is just a power-law superconductor, with suppressed fermion excitation. 
%In this work we focus on the pure and dephased Chern insulator, which corresponds to the Laughlin wave function with $m = 1$. But discussions in this section should apply to fractional Chern insulators with larger $m$. 

%But we expect that it is the amplitude of the strange correlator that is being suppressed by $g$, rather than the scaling dimension. The reason is that, for a 

\section{ Finite dephasing}

\begin{center}
\begin{figure}[h]
\begin{tikzpicture}
\node[anchor=north east] at (1.87,3.5) {\includegraphics[width=3in]{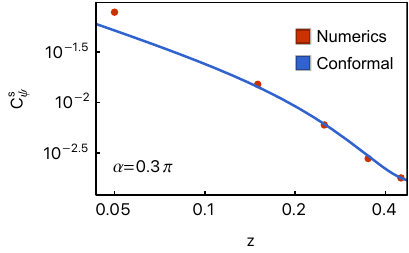}};
\node[black] at (-5.5,3.5) {(a)};
\node[anchor=north east] at (2,-1.5) {\includegraphics[width=3in]{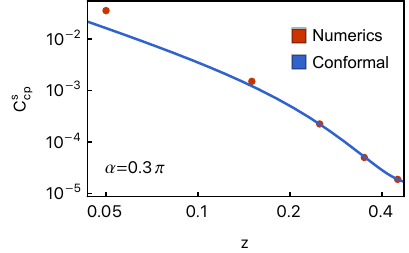}};
\node[black] at (-5.5,-1.5) {(b)};
\end{tikzpicture}
\caption{Single-particle $C_\psi^s$ (a) and Cooper-pair $C_{\text{cp}}^s$ (b) strange correlator as a function of $z = x/L_x$ for a finite decoherence rate $\alpha = 0.3\pi$. The numerical results (red dots) are obtained in the system of size $(L_x,L_y) = (20,8)$ with anti-periodic boundary conditions in both directions. The results agree with the correlation function of a CFT with Luttinger parameter $K = 1$, with partition function Eq.~\ref{AA} (blue curve).
} \label{conformal}
\end{figure}
\end{center}

For finite $g$, the physics should interpolate between the pure state and the Gutzwiller wave function discussed in the previous subsection. As was discussed in Ref.~\cite{cherndecohere}, the dephasing keeps the strong U(1) symmetry, meaning within the bosonized field theory, the strange correlators cannot be short-ranged at finite $g$. The scaling dimensions of all strange correlators are expected to depend only on the same Luttinger parameter of the $c_{\rm eff} = 1$ fidelity-CFT. The theoretical discussions in the previous section indicate that the Cooper pair strange correlator should have the same scaling dimension in the $g = 0$ and $g = \infty$ limit for the trivial-Chern fidelity-CFT, which suggests that the scaling dimensions of all strange correlators should at most weakly depend on $g$, while the amplitude of the strange correlators can vary more strongly with $g$. 

The single particle and Cooper pair strange correlators at different dephasing strengths are plotted in Fig.~\ref{SC}. The correlation functions with nonzero $g$ are calculated by determinant Quantum Monte Carlo, and $\alpha$ is the effective coupling to the Hubbard-Stratanovich field, which is a function of the decoherence strength $g$. Explicitly, $\cos(\alpha) = e^{-g/2}$. As expected, the single particle and Cooper pair strange correlator show opposite trend under dephasing: the Cooper pair strange correlator is enhanced, while the single particle strange correlator is suppressed by dephasing. Note that the strange correlator derived from the pure boson wave function in the infinite $g$ limit has the same scaling as the Cooper pair correlation of a free Dirac fermion, and the same as the Cooper pair strange correlator of the pure Chern insulator. Hence it is the amplitude of the correlators that are suppressed with $g$, while the scaling dimensions remain largely unchanged. These expectations are consistent with the numerical calculation of the strange correlators in Fig.~\ref{SC}. 

We have also compared the computed Cooper pair strange correlator with the correlation function in the compact boson CFT on the torus~\cite{francesco2012conformal}, after combining different twisted sectors as Eq.~\ref{AA}. The numerical result is consistent with the CFT correlation function as shown in Fig.~\ref{conformal}. 

\begin{center}
\begin{figure}[h]
\includegraphics[width=0.38\textwidth]{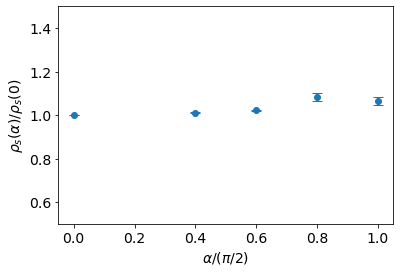}
\caption{The extracted stiffness of the trivial-Chern fidelity-CFT as a function of dephasing strength $\alpha$. The stiffness does not strongly depend on $\alpha$, consistent with the theoretical discussions.} \label{stiffness}
\end{figure}
\end{center}

In Fig.~\ref{stiffness}, we find that the stiffness does not strongly depend on $g$. It has a tendency to slightly increase with $g$ on finite system size, but we find this effect becomes smaller as the system size increases.
%, suggesting that the stiffness is exactly constant in the thermodynamic limit. 
This suggests that the Luttinger parameter $K$ of the fidelity-CFT does not strongly depend on $g$ either. This result is consistent with the behavior of the strange correlators, which are in agreement with the predictions of the $K=1$ compact boson CFT for all $g$, albeit the change of amplitude. This suggests that though dephasing leads to interaction on the fidelity-CFT, and it significantly changes the ordinary Cooper pair correlation function, it does not significantly renormalize the Luttinger parameter of the field theory of the fidelity-CFT. In fact, when $g$ is much smaller than $1$, we can project the density dephasing to the modes localized at the temporal interface. The zero modes of $\psi$ and $\bar{\psi}$ trapped at the temporal interface are eigenstates of $\gamma_0$ with opposite eigenvalues, and the effective interaction caused by dephasing does not overlap with the density of zero modes in the infrared limit. With stronger $g$, the effective interaction caused by dephasing can mix the interface modes with the bulk states, causing more complicated effects. 

\as{There is another argument that indicates that the Luttinger parameter $K$ is not renormalized for all finite $g$. Let us take the reference state $\rho_{tr}$ as a direct product state which is the eigenstate of the local density operator, as expected deep in the trivial insulator phase. The decohered doubled state $|\rho_{c}^g\rrangle$ can be viewed as the solution of the differential equation $\frac{\partial |\rho_c^g\rrangle} {\partial g} = \sum_{k} g \sigma_{i,1}^z \sigma_{i,2}^z |\rho_c^g\rrangle$. We can plug this form into Eq. \ref{corr_eqns} and commute the fermion operators in the correlation function past the $\sigma_{i,a}^z$} operators to find a simple differential equation for $C_\psi^s(i-j)$. If we further assume translational invariance of $\rho_{tr}$, then this equation takes the simple form %The decohered doubled state $\rho_c^g$ can be viewed as the solution to a finite-time Lindbladian evolution, with jump operators given by the local density. This perspective implies that the strange correlator at finite $g$ satisfies the differential equation, which is analogous to evolution of correlation matrix with the Lindbladian dynamics~\cite{dolgirev2020,turkeshi2021,kiely2025phasetransitiontopologicalindex}}
\beqn
\partial_g C_\psi^s(i-j) = - 4 g C_\psi^s(i-j)
\eeqn

\noindent %which forbids renormalization of the scaling dimension. 
which holds for $i \neq j$. \cx{We can take an power-law ansatz of the strange correlator $C_\psi^s(i-j) = c/|i - j|^{2\Delta_\psi}$, and demonstrate that only the amplitude $c$ will evolve under dephasing strength $g$, while $\Delta_\psi$ remains a constant under $g$, which implies that the Luttinger parameter of the effective CFT remains unchanged under dephasing.}
%where the sum runs over lattice sites. The scaling behavior of $C_
%\psi^s$ should not evolve under this differential equation.}

Another question worth asking is whether the power-law behavior in the ordinary Cooper pair correlation emerges only at $g \ra \infty$, or at finite critical $g_c$. For $2d$ quantum states a transition can indeed be driven by finite dephasing, like the decodability transition of the $2d$ toric code. As we discussed, the doubled state in the limit $g \ra \infty$ is a Gutzwiller wave function. One potential interpretation of the Gutzwiller projection is by viewing its effect as coupling the system to a dynamical gauge field. The emergent superconductivity is the condensate of the gauge flux which traps charge-$2$ through the quantum spin Hall effect in the doubled space~\cite{ransu2} (though this picture would lead to a true long range rather than power-law superconductor). This picture suggests that the emergent superconductivity only occurs at $g \ra \infty$, as that is the case where the gauge constraint is strictly enforced.  

%In the Laughlin wave function and Coulomb gas representation of the Chern insulator, turning on dephasing will lead to short-range attractive interaction between $z_i$ and $w_i$ ( Eq.~\ref{laughlinSF}). However, this attractive interaction is extremely local, and we do not expect it to modify the long distance behavior of the Coulomb gas, unless the local attraction is infinitely strong, which corresponds to the limit of $g \ra \infty$. 

%maps to a density-density interaction between the doubled states; and under a space-time rotation, the Dirac fermion density $\bar{\psi}\gamma^0 \psi$ becomes the axial mass term $\bar{\psi}\gamma^5 \psi$. 

\section{Conclusion and discussion}

In this work we demonstrated that the fidelity and 2nd relative R\'{e}nyi entropy between two states with different Chern numbers map nicely to a $(2+0)d$ CFT living on the temporal interface. The dephasing maps to certain interaction of the CFT. In the infinite dephasing limit, the doubled state of the dephased Chern insulator becomes a Gutzwiller projected quantum spin Hall insulator, which is expected to be a power-law superconductor based on analysis of the Laughlin form of the wave function. Our numerical results are qualitatively compatible with theoretical predictions. 

In this work we focused on the effect of density dephasing on Chern insulators, which is the most physically relevant decoherence channel. We found that dephasing does not strongly renormalize the Luttinger parameter of the field theory of the fidelity-CFT living on the temporal interface. If the Kraus operators involve the current operators, it is expected to more obviously renormalize the Luttinger parameter~\cite{cherndecohere}. We leave this to future numerical study. 

We can also consider related problems of dephased fractional Chern insulator, and more generally dephased fractional topological insulators. The fractional Chern insulators feature rich physics, including both 't Hooft anomaly and anyons. These two phenomena may undergo different evolutions under dephasing, leading to a richer phase diagram. We will also leave this exploration to future work. 

\cx{The strange correlator was designed as a tool to diagnose the anomaly of the system at the temporal boundary, which arises from the topological effect in the space-time bulk. Here we briefly discuss its application in diagnosing topological orders. Let us assume that our target system is a chiral topological order with a Chern-Simons (CS) theory description. Due to the fractionalization of physical fermion, the single particle strange correlator against a trivial reference state (which maps to the Green's function at the boundary of the system) decays as $1/r^{k}$, where $k$ is the level of the CS term. Therefore the single particle strange correlator could potentially serve as a rather efficient tool of diagnosing chiral topological order. For interacting nonchiral topological order, the behaviors of strange correlator can be complicated by the Luttinger parameter, therefore it would take more efforts to fully identify a nonchiral topological order. We leave the detailed discussion as well as numerical test to future study. }

%\as{Lastly, in this work, we considered the problem of strange correlators of a dephased Chern insulator, which corresponds to perturbations on the boundary of an abelian Chern-Simons theory. For any abelian topological order described by a $K$-matrix, in principle a Laughlin-like wavefunction $\ket{\phi_K}$ can be constructed. Thus, if we are given a wavefunction corresponding to some ``target" topological order $\ket{\psi_t}$, we can compute the strange correlators between the target topological order and various wavefunctions for Abelian topological orders $\cZ_K = |\langle \psi_t | \phi_K \rangle|^2$, allowing us to probe the physics of domain walls between the target topological order and the $K$-matrix Chern-Simons theory, which would allow us to distill more useful information of the target topological order. For example, if any gapped domain wall is allowed between the two orders, then generic weak dephasing would flow under RG to gap out $\cZ_K$, leading the strange correlator to be short ranged. The existence of such a gapped domain wall implies the existence of a tunneling matrix between the anyon types in the two topological orders satisfying various consistency conditions, which strongly constrains the target order \cite{Lan_2015}.}

{\it --- Acknowledgment:} The authors thank Matthew P. A. Fisher and Wenjie Ji for very helpful discussions. A.S., N.M.J., and C.X. are supported by the Simons Foundation International through the Simons Investigator grant. Y.B. and T.K. are supported in part by grant NSF PHY-2309135 to the Kavli Institute for Theoretical Physics (KITP). Y.B. is supported in part by the Gordon and Betty Moore Foundation Grant No. GBMF7392 to the KITP. T.K. is supported in part by the Gordon and Betty Moore Foundation Grant No. GBMF8690 to the KITP.

\bibliography{big}

\end{document}